# Exact Solutions for Spin Conserving Models and the Wigner–Araki–Yanase Theorem

Michael Steiner[1] and Ronald Rendell[2]
[1,2]Inspire Institute, Alexandria VA

Abstract

The Wigner-Araki-Yanase (WAY) theorem is a well-known theorem regarding limitations of quantum measurement in the presence of additive conservation laws. Under the assumptions of the von Neumann measurement model, for which the system conserved quantity $L_S$ is bounded, given a conserved total additive system plus apparatus quantity $L_{SA}$, the measurement operator $E_S$ must commute with $L_S$. Prior proofs have exploited the properties of unitary evolution constrained by momentum conserving operations that tend to obscure the physical nature of the WAY theorem and as well lead to bounds on performance. As it is generally agreed that momentum is always exactly conserved in measurement, we instead develop a general angular momentum conserving model of measurement. This model is shown to lead to a simple explanation of the major implications of the WAY theorem and provides exact results of the effects of measurement based on the apparatus model. This is shown by both tracing the apparatus from the density matrix and also via a system-only channel model based on Kraus operators.

## 1. Introduction

Consider a spin initially prepared spin up relative to the *z*-direction, denoted $|\uparrow\rangle_z$ that enters a Stern-Gerlach device oriented to obtain measurements in the *x*-direction (or similar device). The initial system state is equivalent to the following superposition of states of spin in the *x*-direction:

$$|\uparrow\rangle_z = \frac{1}{\sqrt{2}}(|\uparrow\rangle_x + |\downarrow\rangle_x). \tag{1}$$

If one were to subject this system to the measurement $\sigma_x$ where $\sigma_x$ is the Pauli spin matrix for a spin measured in the *x*-direction, one can ask if angular momentum is conserved. Let $L_{z,SA}$ be the total angular momentum in the *z*-direction of the composite system plus apparatus. $L_{z,SA}$ is found by computing the additive quantity $L_{z,SA} \equiv L_{z,S} \otimes I + I \otimes L_{z,A}$ where $L_{z,S}$ is the system angular momentum in the $z$-direction and $L_{z,A}$ is the apparatus angular momentum in the $z$-direction. Let us denote $L_{i,z,SA}$ as the initial angular momentum of the system plus apparatus in the $z$-direction and $L_{f,z,SA}$ as the final angular momentum after a measurement is made. It is assumed that the initial apparatus angular momenta in the *x,y,* and $z$-directions are zero and that for simplicity only spin angular momentum is assumed to be present throughout the evolution of the particle in the measurement device.

Given the initial system state of $|\uparrow\rangle_z$, the initial expected values of system and apparatus z-directed angular momenta are given by $E(L_z \otimes I) = \hbar/2$ and $E(I \otimes L_z) = 0$, hence $E(L_{i,z,SA}) = \hbar/2$. Consider a measurement in the $x$-direction with $O_S$=$\sigma_x$ and suppose that spin up is observed. Then the system obtains spin $|\uparrow\rangle_x$ as a result of the measurement. The initial composite spin angular momentum in the $x$ direction is $L_{x,SA} \equiv L_{x,S} \otimes I + I \otimes L_{x,A}$ and $E(L_{x,SA}) = 0$ since $E(L_{x,S} \otimes I) = 0$ and $E(I \otimes L_{x,A}) = 0$. In order for angular momentum to be conserved in the $x$-direction, the spin of the apparatus must increase by the spin of $|\downarrow\rangle_x$. Hence the spin of system-apparatus becomes $|\uparrow\rangle_x \otimes |\psi_{app}{+}\downarrow\rangle_x$ where $|\psi_{app}{+}\downarrow\rangle_x$ represents the state with the addition of spin angular momentum of $|\downarrow\rangle_x$ to the initial state of the apparatus $|\psi_{app}\rangle$. At the final time it is seen that $E(L_{x,S} \otimes I) = \hbar/2$ and $E(I \otimes L_{x,A}) = -\hbar/2$, hence $E(L_{x,SA}) = 0$. Yet the total angular momentum in the $z$-direction is not conserved: initially $E(L_{i,z,SA}) = \hbar/2$, and after the measurement it is seen that $E(L_{f,Z,T}) = 0$ since both $E(L_{z,S} \otimes I) = 0$ and $E(I \otimes L_{z,A}) = 0$, whereas $E(L_{i,z,SA}) = \hbar/2$.

The example just considered illustrates the Wigner–Araki–Yanase (WAY) theorem. The WAY theorem makes the following assumptions: 1) that $U$ is derived from the particular unitary interaction proposed by von Neumann in Eqn. (2), 2) that $U$ and $L_T$ commute and $L_T$ is a conserved quantity during unitary interaction, 3) $L_S$ is bounded, and 4) $L_T$ is conserved as the result of measuring $E_S$. The WAY theorem concludes that $E_S$ must commute with $L_S$. Wigner in [1], which was generalized in [2], showed that if a quantity such as momentum is conserved during measurement, then only approximate measurement is possible when the measurement does not commute with the conserved quantity. A large apparatus was shown to be necessary in order to provide good approximate measurement.

The WAY theorem [1] was generalized by several authors including Araki-Yanase [2], Yanase [3], Shimony and Stein [4], Ghirardi et al [5], [6] and Ozawa [7]. A common manner of proof of the WAY theorem is to utilize the properties of unitary evolution along with the constraints of conservation to prove that when measuring an operator $M$, then $M$ must commute with $L$. For example, Eqns. (8)-(17) from [1] exploit the properties of unitary evolution to show that the device must measure in an indeterminate manner related to the inverse of a quantity related to the size of the conserved quantity inherent in the apparatus. Similarly, Eqns. (2.3)-(2.14) of Yanase [3] are derived from the properties of unitary evolution and conservation to eventually find a bound on the probability of unsuccessful measurement. In Araki-Yanase [2], although a proof is given that shows that $M$ must commute with $L$ under a conservation constraint, the result does not provide an intuitive reason as to why there should be failure in such measurements. Similarly, work by Ghirardi et al [5], [6] derives a lower bound on the probability of unsuccessful measurement using unitary and conservation constraints in Eqns. (3.4)-(3.16) of their paper [5].

In this paper, instead of proving as in the WAY theorem that momentum conservation implies that $E_S$ must commute with $L_S$, we consider a general class of momentum and energy conserving models of measurement. This is a reasonable approach, because it is generally agreed that momentum and energy are not approximately or even asymptotically conserved, but rather are exactly conserved. This class of models is useful for several reasons. Firstly, it will be seen to be sufficiently general to illustrate precisely why measurement must introduce errors when one is attempting to measure spin in the $x$-direction when a spin is prepared initially in a spin up or down state in the $z$-direction. Secondly, no complicated arguments using the properties of unitary evolution and conservation are utilized. Thirdly, a simple bound can be derived which illustrates why the apparatus must have a substantial uncertainty in order for measurement to occur with good approximation. Fourthly, exact solutions of how the system state is affected by measurement can be obtained with a model of how the uncertainty in the SG device is manifested.

In Sec. 2, a straightforward energy and momentum conserving model is presented to measure exactly the case when a spin eigenstate initially prepared in the $z$-direction is subject to a Stern-Gerlach measurement in the $x$-direction. It will be seen that although the statistics of the measurement are correct for this prepared state, when the initial state is instead an eigenstate in the $x$-direction, errors must occur. The errors are seen to reduce as a function of the apparatus initial uncertainty of spin angular momentum. Using a model of the uncertainty in the SG device, we show in Sec. 3 how to derive exact solutions of how the system state is affected by measurement. Conclusions are drawn in Sec. 4.

## 2. Energy and Momentum Conserving Models

Consider a system and apparatus with states in the respective Hilbert spaces $\mathcal{H}_S$ and $\mathcal{H}_A$. The theorem often referred to as the WAY theorem considers the measurement of a sharp system observable $E_S$ with discrete eigenvalues $\lambda_i$ and corresponding eigenvectors $e_{i,j}$

$$E_S|e_{i,j}\rangle = \lambda_i|e_{i,j}\rangle, \forall i, j.$$

with orthogonal eigenvectors i.e. $\langle e_{i,j}|e_{k,l}\rangle = \delta_{i,k}\delta_{j,l}$, and $\delta_{a,b} = 1$ if $a = b$, $\delta_{a,b} = 0$ if $a \neq b$. The Hamiltonian is such that at $t = 0$ the system and apparatus are non-interacting and that at some intermediate time the system interacts with the apparatus, followed by a gradual decrease in the interaction between system and apparatus as they become separated until at long times, which we assume is at $t = T$, no interaction exists. Wigner originally assumed a measurement model proposed by von Neumann. The von Neumann measurement model assumes that if the initial state of the apparatus is $|\varphi_A\rangle$ and the system is initially in any eigenstate of an associated observable $S$ with the $i$ th outcome $|e_{i,j}\rangle$, the unitary evolution of the initial state $|e_{i,j}\rangle\otimes|\varphi_A\rangle$ results in

$$U\,|e_{i,j}\rangle\otimes|\varphi_A\rangle = \sum_j a_{i,j}\,|e_{i,j}\rangle\otimes|\varphi_{A,i,j}\rangle \tag{2}$$

whereby $|\varphi_{A,i,j}\rangle$ represents multiple apparatus states which point to the eigenvalue $\lambda_i$ having been measured, or the $i$ th outcome having occurred. Essentially if one starts in an eigenstate of the observable being measured, the unitary evolution results in the apparatus pointing to the correct outcome with probability unity. The quantum states representing the apparatus are orthogonal whenever different pointer states result, i.e. $\langle\varphi_{A,i,j}\,|\varphi_{A,i,j'}\rangle = 0$ when $j \neq j'$. It is seen that one can associate a pointer apparatus observable represented by a Hermitian operator $E_A$ that corresponds to the system observable $E_S$ via Eqn. (2). A total additive conserved quantity $L_T$ is defined as

$$L_T = L_S \otimes I + I \otimes L_A$$

where $L_S$ denotes the quantity in the system and $L_A$ in the apparatus. Note that $L_T$ is additive in the sense that if $L_S$ and $L_A$ were to measure a similar quantity, for example spin, then $L_T$ would be a spin vector of the total spin in the $x, y, z$ directions consisting of the sum of the spin of the system and the spin of the apparatus direction.

We now consider a general class of models that always conserve the total quantity $L_T$. For simplicity, we assume as in the example in Sec. 1 that we will be measuring using a Stern-Gerlach or similar device that is oriented in the *x*-direction for an initial spin $|\uparrow\rangle_{z,S}$ that is spin-up in the *z*-direction. We desire to conserve spin angular momentum both in the *z*-direction and *x*-direction. We first utilize a unitary operator $U$ of Eqn. (2) that maps as follows:

$$\begin{aligned}
&|\uparrow\rangle_{z,S}\otimes|0\rangle_{x,A}\otimes|0\rangle_{z,A}\otimes|0\rangle_{p,A} \\
&\qquad\to (|\uparrow\rangle_{x,S}\otimes|{+\downarrow}\rangle_{x,A}\otimes|{+\uparrow}\rangle_{z,A}\otimes|\chi_{\uparrow,x}\rangle_{p,A} \\
&\qquad\quad + |\downarrow\rangle_{x,S}\otimes|{+\uparrow}\rangle_{x,A}\otimes|{+\uparrow}\rangle_{z,A}\otimes|\chi_{\downarrow,x}\rangle_{p,A})/\sqrt{2} \\
&|\downarrow\rangle_{z,S}\otimes|0\rangle_{x,A}\otimes|0\rangle_{z,A}\otimes|0\rangle_{p,A} \\
&\qquad\to (|\uparrow\rangle_{x,S}\otimes|{+\downarrow}\rangle_{x,A}\otimes|{+\downarrow}\rangle_{z,A}\otimes|\chi_{\uparrow,x}\rangle_{p,A} \\
&\qquad\quad - |\downarrow\rangle_{x,S}\otimes|{+\uparrow}\rangle_{x,A}\otimes|{+\downarrow}\rangle_{z,A}\otimes|\chi_{\downarrow,x}\rangle_{p,A})/\sqrt{2}.
\end{aligned} \tag{3}$$

where the state $|\uparrow\rangle_{z,S}\otimes|0\rangle_{x,A}\otimes|0\rangle_{z,A}\otimes|0\rangle_{p,A}$ represents an initial system state $|\uparrow\rangle_{z,S}$ of spin up in the *z*-direction, and the remaining three components $|0\rangle_{x,A}\otimes|0\rangle_{z,A}\otimes|0\rangle_{p,A}$ represent apparatus angular momentum states in the *x*-direction, *z*-direction, and $|0\rangle_{p,A}$ is the initial pointer state of the apparatus. Although the constraints of Eqn. (2) only result in a rank-2 matrix, one can easily complete the matrix so that $U$ is unitary—the unitary that results for the specific model of Sec. 3.1 is shown in Appendix A. This state is mapped to the sum of two states, where the first state, $|\uparrow\rangle_{x,S}\otimes|{+\downarrow}\rangle_{x,A}\otimes|{+\uparrow}\rangle_{z,A}\otimes|\chi_{\uparrow,x}\rangle_{p,A}$ represents the system in spin-up in the *x*-direction, the state $|{+\downarrow}\rangle_{x,A}$ represents the addition of spin-down in the *x*-direction that has been transferred to the apparatus from the initial state $|0\rangle_{x,A}$, $|{+\uparrow}\rangle_{z,A}$ similarly represents the addition of spin-up in the *z*-direction that has been transferred to the apparatus from the initial apparatus state $|0\rangle_{z,A}$, and $|\chi_{\uparrow,x}\rangle_{p,A}$ represents a pointer state of the Stern-Gerlach device that represents pointing to a system state that has passed through the Stern-Gerlach device as would a spin-up system state in the *x*-direction.

Note that Eqn. (3) can be used to represent a rather general class of models because the initial apparatus state $|0\rangle_{x,A}\otimes|0\rangle_{z,A}\otimes|0\rangle_{p,A}$ can be chosen arbitrarily. Also note that this unitary evolution operator conserves *z*-direction spin exactly. It does this by transferring the initial system spin $|\uparrow\rangle_{z,S}$ into the apparatus $|{+\uparrow}\rangle_{z,A}$. In order to conserve *x*-direction spin exactly consider the following. Since $|\uparrow\rangle_{z,S}$ can also be represented by an equal superposition of spin up and down in the *x*-direction, i.e. $|\uparrow\rangle_{z,S} = (|\uparrow\rangle_{x,S} + |\downarrow\rangle_{x,S})/\sqrt{2}$, we map $|\uparrow\rangle_{z,S}\otimes|0\rangle_{x,A}\otimes|0\rangle_{z,A}\otimes|0\rangle_{p,A}$ into an equal superposition of $|\uparrow\rangle_{x,S}\otimes|{+\downarrow}\rangle_{x,A}\otimes|{+\uparrow}\rangle_{z,A}\otimes|\chi_{\downarrow,x}\rangle_{p,A}$ and $|\downarrow\rangle_{x,S}\otimes|{+\uparrow}\rangle_{x,A}\otimes|{+\uparrow}\rangle_{z,A}\otimes|\chi_{\downarrow,x}\rangle_{p,A}$. In this case, the term containing $|\uparrow\rangle_{x,S}$ is paired with equal and opposite momentum in the apparatus $|{+\downarrow}\rangle_{x,A}$ so that detection

of this term (or the term with $|\downarrow\rangle_{x,S}$) continues to exactly conserve spin angular momentum. Such a class of spin conserving models appear to be sufficiently general to allow further study.

With a class of spin-conserving models having been defined, we can now investigate precisely what happens for different potential inputs. First, let us consider an initial system state of $|\uparrow\rangle_{z,S}$. Unitary evolution followed by a projective measurement of the Pauli matrix $\sigma_x$ on the system results in the initial state of $|\uparrow\rangle_{z,S}\otimes|0\rangle_{x,A}\otimes|0\rangle_{z,A}$ evolving correctly to either $|\uparrow\rangle_{x,S}\otimes|+\downarrow\rangle_{x,A}\otimes|+\uparrow\rangle_{z,A}\otimes|\chi_{\downarrow,x}\rangle_{p,A}$ or $|\downarrow\rangle_{x,S}\otimes|+\uparrow\rangle_{x,A}\otimes|+\uparrow\rangle_{z,A}\otimes|\chi_{\downarrow,x}\rangle_{p,A}$ with 50% probability. In both cases, both $x$ and $z$ direction angular momenta are exactly conserved. One might expect that mathematically, this is equivalent to a rank-1 projection into either $|\uparrow\rangle_{x,S}$ or $|\downarrow\rangle_{x,S}$. Interestingly, this is not strictly correct due to pairing of $|\uparrow\rangle_{x,S}$ with $|+\downarrow\rangle_{x,A}$ versus the pairing of $|\downarrow\rangle_{x,S}$ with $|+\uparrow\rangle_{x,A}$; this is further discussed in Sec. 3. In any case, with input $|\uparrow\rangle_{z,S}$, it is expected that when measuring in the $x$-direction, the correct probabilities result.

We desire that the SG apparatus should not only function if $|\uparrow\rangle_{z,S}$ is input but should still function if $|\uparrow\rangle_{x,S}$ or $|\downarrow\rangle_{x,S}$ is input. In this case we desire that both $|\uparrow\rangle_{x,S}\otimes|0\rangle_{p,A}\rightarrow|\uparrow\rangle_{x,S}\otimes|\chi_{\uparrow,x}\rangle_{p,A}$ and $|\downarrow\rangle_{x,S}\otimes|0\rangle_{p,A}\rightarrow|\downarrow\rangle_{x,S}\otimes|\chi_{\downarrow,x}\rangle_{p,A}$. This is an assumption in the vN model of measurement. Now, if this is met, then the WAY theorem should be valid and we should not be able to conserve momentum. On the other hand, in our model we do conserve momentum, so the possibility that spin in the $x$-direction is not measured correctly needs to be considered. One can actually derive what will happen with these inputs from Eqns. (3) above using $|\uparrow\rangle_{x,S}=\frac{|\uparrow\rangle_{z,S}+|\downarrow\rangle_{z,S}}{\sqrt{2}}$. By linearity, this gives a superposition of Eqns. (3), i.e. $|\uparrow\rangle_{z,S}\otimes|0\rangle_{x,A}\otimes|0\rangle_{z,A}\otimes|0\rangle_{p,A}=(|\uparrow\rangle_{x,S}\otimes|0\rangle_{x,A}\otimes|0\rangle_{z,A}\otimes|0\rangle_{p,A}+|\downarrow\rangle_{x,S}\otimes|0\rangle_{x,A}\otimes|0\rangle_{z,A}\otimes|0\rangle_{p,A})/\sqrt{2}$ evolves to

$$\begin{aligned}(&|\uparrow\rangle_{x,S}\otimes|+\downarrow\rangle_{x,A}\otimes|+\uparrow\rangle_{z,A}\otimes|\chi_{\uparrow,x}\rangle_{p,A}+\\ &|\downarrow\rangle_{x,S}\otimes|+\uparrow\rangle_{x,A}\otimes|+\uparrow\rangle_{z,A}\otimes|\chi_{\downarrow,x}\rangle_{p,A}+\\ &|\uparrow\rangle_{x,S}\otimes|+\downarrow\rangle_{x,A}\otimes|+\downarrow\rangle_{z,A}\otimes|\chi_{\uparrow,x}\rangle_{p,A}-\\ &|\downarrow\rangle_{x,S}\otimes|+\uparrow\rangle_{x,A}\otimes|+\downarrow\rangle_{z,A}\otimes|\chi_{\downarrow,x}\rangle_{p,A})/2\end{aligned} \tag{4}$$

Now, the first and third terms reduce to $|\uparrow\rangle_{x,S}\otimes(|+\downarrow\rangle_{x,A}\otimes|+\uparrow\rangle_{z,A}+|+\downarrow\rangle_{x,A}\otimes|+\downarrow\rangle_{z,A})\otimes|\chi_{\uparrow,x}\rangle_{p,A}$ and these terms would correctly readout $|\uparrow\rangle_{x,S}$ while conserving momentum. However, the second and fourth terms need to cancel for this to happen. The second plus fourth terms are given by

$$(|\downarrow\rangle_{x,S}\otimes|+\uparrow\rangle_{x,A}\otimes|+\uparrow\rangle_{z,A}\otimes|\chi_{\downarrow,x}\rangle_{p,A}-|\downarrow\rangle_{x,S}\otimes|+\uparrow\rangle_{x,A}\otimes|+\downarrow\rangle_{z,A}\otimes|\chi_{\downarrow,x}\rangle_{p,A}.$$

The problem is that the above second and fourth terms do not generally cancel. Note that if the initial angular momentum of the device in the $x$ and $z$ direction have a substantial uncertainty in angular momentum, then one can consider $|+\downarrow\rangle_{z,A}=\sqrt{\alpha}|+\uparrow\rangle_{z,A}+\sqrt{1-\alpha}|+\uparrow_{\perp}\rangle_{z,A}$, where $|+\uparrow_{\perp}\rangle_{z,A}$ is a state orthogonal to $|+\uparrow\rangle_{z,A}$. Noting that ${}_{z,A}\langle+\uparrow|+\downarrow\rangle_{z,A}=\sqrt{\alpha}$ , then

$$|\downarrow\rangle_{x,S}\otimes|+\uparrow\rangle_{x,A}\otimes|+\uparrow\rangle_{z,A}\otimes|\chi_{\downarrow,x}\rangle_{p,A}-|\downarrow\rangle_{x,S}\otimes|+\uparrow\rangle_{x,A}\otimes|+\downarrow\rangle_{z,A}\otimes|\chi_{\downarrow,x}\rangle_{p,A}$$

$$= |\downarrow\rangle_{x,S}\otimes|+\uparrow\rangle_{x,A}\otimes|+\uparrow\rangle_{z,A}\otimes|\chi_{\downarrow,x}\rangle_{p,A} - |\downarrow\rangle_{x,S}\otimes|+\uparrow\rangle_{x,A}\otimes(\sqrt{\alpha}|+\uparrow\rangle_{z,A} - \sqrt{1-\alpha}|+\uparrow_{\perp}\rangle_{z,A})\otimes|\chi_{\downarrow,x}\rangle_{p,A}$$
$$= (1-\sqrt{\alpha})|\downarrow\rangle_{x,S}\otimes|+\uparrow\rangle_{x,A}\otimes|+\uparrow\rangle_{z,A}\otimes|\chi_{\downarrow,x}\rangle_{p,A} - \sqrt{1-\alpha}|\downarrow\rangle_{x,S}\otimes|+\uparrow\rangle_{x,A}\otimes|+\uparrow_{\perp}\rangle_{z,A}\otimes|\chi_{\downarrow,x}\rangle_{p,A}$$

If one expects $\alpha \approx 1$, then this can be approximated as

$$-\sqrt{1-\alpha}|\downarrow\rangle_{x,S}\otimes|+\uparrow\rangle_{x,A}\otimes|+\uparrow_{\perp}\rangle_{z,A}\otimes|\chi_{\downarrow,x}\rangle_{p,A}$$

which indicates $|\downarrow\rangle_{x,S}$ will be (incorrectly) read-out with error rate probability $\frac{(-\sqrt{1-\alpha})^2}{4} = (1-\alpha)/4$ when $|\uparrow\rangle_{x,S}$ is input. Now, so long as ${}_{z,A}\langle+\uparrow|+\downarrow\rangle_{z,A} \cong 1$, such cancellation will occur. However, this requires significant uncertainty in the SG device that generates the magnetic field used to separate the particles. Using our measurement model from Eqn. (4), one can find exact solutions for how the uncertainty in the SG device effects the system state. This is further illustrated in the next section.

## 3. Exact Solutions

In this section we will illustrate how one can obtain an exact solution for the impact that the SG device will have on the system state. In order to determine how the uncertainty in the SG device effects the system state, one needs to consider how the apparatus state, which is initially $|0\rangle_{x,A}\otimes|0\rangle_{z,A}$, forms apparatus states such as $|+\uparrow\rangle_{x,A}$ and $|+\uparrow\rangle_{z,A}$.

In order that apparatus states such as $|+\uparrow\rangle_{x,A}$ and $|+\uparrow\rangle_{z,A}$ are formed so that spin angular momentum is conserved, a model of the spin uncertainty within the apparatus is needed. A simple model to consider initially is found by specifying parametrically the states as follows: $|+\uparrow\rangle_{x,A} = (0\ \sqrt{\beta_x}\ \sqrt{1-\beta_x}\ \ 0)$, $|+\downarrow\rangle_{x,A} = (0\ \sqrt{\beta_x}\ 0\ \sqrt{1-\beta_x})$ and similarly $|+\uparrow\rangle_{z,A} = (0\ \sqrt{\beta_z}\ \sqrt{1-\beta_z}\ \ 0)$, $|+\downarrow\rangle_{z,A} = (0\ \sqrt{\beta_z}\ 0\ \sqrt{1-\beta_z})$. Essentially, the states $|+\uparrow\rangle_{x,A}$ and, $|+\downarrow\rangle_{x,A}$ (and similarly the states in the *z*-direction) are generally non-orthogonal with correlation ${}_{x,A}\langle+\downarrow|+\uparrow\rangle_{x,A} = \beta_x$, ${}_{z,A}\langle+\downarrow|+\uparrow\rangle_{z,A} = \beta_z$. It is assumed that $|+\uparrow\rangle_{x,A}$ has one additional spin-up angular momentum in the *x*-direction component over $|0\rangle_{x,A}$. Similarly in order to have one additional spin-down angular momentum in the *x*-direction component between $|0\rangle_{x,A}$ and $|+\downarrow\rangle_{x,A}$, there must be at least one additional component which is why the fourth component is included. Similar relationships hold for spin in the *z*-direction. The unitary matrix that results with this model is shown in Appendix A.

### 3.1 Solution by Tracing Apparatus

With this model, one can now consider the effect of *U* on several initial states by tracing the apparatus from the overall composite system-apparatus. Consider first the evolution of the initial system state $|\uparrow\rangle_{z,S}$ which in vector notation will be denoted as $|\uparrow\rangle_{z,S} = \binom{1}{0}$. Defining $|\psi_{f,\uparrow,z}\rangle \equiv U|\uparrow\rangle_{z,S}\otimes|0\rangle_{x,A}\otimes|0\rangle_{z,A}\otimes|0\rangle_{p,A}$, we desire to know the effect of the apparatus internal states $x, A$ and $z, A$ which are characterized by $\beta_x$ and $\beta_z$ respectively on the evolution. Hence, we desire to find the density matrix of the system after unitary evolution of the initial state by tracing out the apparatus states, i.e. the final system density matrix is computed by $\rho_{f,S} \equiv \mathrm{Tr}_{p,A}|\psi_{f,\uparrow,z}\rangle\langle\psi_{f,\uparrow,z}|$. The result for an initial state of $|\uparrow\rangle_{z,S}$ is $\rho_{f,S} = |\uparrow\rangle_{z,S\,z,S}\langle\uparrow|$ ; hence there is no change for $|\uparrow\rangle_{z,S}$ and also no change is found for the input $|\downarrow\rangle_{z,S}$. Now, let us consider the input $|\uparrow\rangle_{x,S}$ which in vector notation is given by $|\uparrow\rangle_{x,S} = \frac{1}{\sqrt{2}}\binom{1}{1}$. In this case, a simple closed form solution is found:

$$\rho_{f,S} = \begin{pmatrix} 1/2 & \beta_z/2 \\ \beta_z/2 & 1/2 \end{pmatrix}. \tag{5}$$

Note that when $\beta_z = 1$, $\rho_{f,S} = |\uparrow\rangle_{x,S}\,{}_{x,S}\langle\uparrow|$ and the input is correctly reproduced. However, for $\beta_z < 1$, it is seen that the off-diagonal elements approach 0 as $\beta_z \to 0$. Hence coherence is lost as a direct linear function of the correlation parameter $\beta_z$ of the apparatus. Similar results are obtained for initial $|\downarrow\rangle_{x,S}$. This equivalently shows that for inputs of eigenstates of $\sigma_x$, such as spin $|\uparrow\rangle_{x,S}$, it is possible that an error of $|\downarrow\rangle_{x,S}$ results as a function of $\beta_z$.

An interesting question is what effect such an apparatus has on a maximally entangled state of two spins, of which one of the spins is input into an apparatus. A maximally entangled state has a reduced density matrix of

$$\rho_S = \begin{pmatrix} 1/2 & 0 \\ 0 & 1/2 \end{pmatrix} = \frac{I}{2}$$

Hence $\rho$ is a 50%-50% mixed state of $|\uparrow\rangle_{z,S}$ and $|\downarrow\rangle_{z,S}$. Since the effect of the unitary operation on either $|\uparrow\rangle_{z,S}$ or $|\downarrow\rangle_{z,S}$ is to leave the state unchanged, one might at first expect then that the identity matrix of the system will be maintained by the unitary and the entanglement be unchanged. In fact, it was found that the partial density matrix of the system does remain unchanged by the unitary and $\rho_{f,S} = \mathrm{Tr}_{p,A} U(\rho_{S,i} \otimes \rho_{A,i})U'$ is given by $\rho_{f,S} = I/2$ when $\rho_{S,i} = I/2$. However, even with such a unitary mapping, it is surprisingly not necessarily the case that the entanglement remains the same. A closed form derivation was performed for $\rho_{f,S} = \mathrm{Tr}_{p,A}(I \otimes U)(\rho_{12,i} \otimes \rho_{A,i})(I \otimes U)'$ where $U$ acts on the second particle of the two-particle state $\rho_{12,i}$ for which

$$\rho_{12,i} = \begin{pmatrix} 0 & 0 & 0 & 0 \\ 0 & 1/2 & 1/2 & 0 \\ 0 & 1/2 & 1/2 & 0 \\ 0 & 0 & 0 & 0 \end{pmatrix}.$$

It is found that

$$\rho_{S,f} = \begin{pmatrix} 0 & 0 & 0 & 0 \\ 0 & 1/2 & \beta_z/2 & 0 \\ 0 & \beta_z/2 & 1/2 & 0 \\ 0 & 0 & 0 & 0 \end{pmatrix}.$$

To see why this is the case, we will now turn to the development of a system-only channel model. With such a model it will become clear as to why the entanglement decreases.

## 3.2 Solution with System Channel Model

In order to better characterize the effect of the apparatus on the system, particularly when one particle of a maximally entangled state is input, a system-only representation using Kraus superoperators is now developed. Given an initial system density matrix $\rho_{S,i}$ and initial apparatus state from Eqn. (4) of $|\psi\rangle_{i,A} = |0\rangle_{x,A} \otimes |0\rangle_{z,A} \otimes |0\rangle_{p,A}$ the initial system-apparatus density matrix is given by $\rho_{S,i} \otimes \rho_{A,i} =$

$\rho_{S,i} \otimes |\psi\rangle_{i,A\,i,A}\langle\psi|$. The final system state after unitary evolution was found in Sec. 3.1 by tracing out the apparatus, i.e. via $\rho_{f,S} = \mathrm{Tr}_{p,A} U(\rho_{S,i} \otimes \rho_{A,i})U'$. However, it is known that a system-only representation via Kraus superoperators $M_i$ can also be found for which

$$\rho_{f,S} = \sum_i M_i \rho_{S,i} M_i'.$$

Preskill has shown that one can obtain these Kraus operators via [8, Eqn. (3.68)]:

$$M_\mu = {}_{p,A}\langle\mu|U|0\rangle_{p,A}$$

where $|\mu\rangle_{p,A}$ is an orthogonal basis for $p, A$ and $U$ is based on Eqn. (2) that is further completed to a unitary matrix. We utilize $|\mu\rangle_{p,A}$ in the natural basis, i.e. for $\mu = 1$, $|1\rangle_{p,A} = (1, 0, .. ,0)$. For our problem, when computing these $M_\mu$, one finds that there are six nonzero $M_\mu$. However, the form of each $M_\mu$ is found to be the same as a different $M_\alpha$ such that the six can be further reduced to the following three $M_\mu$:

$$\begin{aligned} M_1 &= \sqrt{\beta_z}\begin{pmatrix} 1 & 0 \\ 0 & 1 \end{pmatrix} \\ M_2 &= \begin{pmatrix} \sqrt{1-\beta_z} & 0 \\ 0 & 0 \end{pmatrix} \\ M_3 &= \begin{pmatrix} 0 & 0 \\ 0 & \sqrt{1-\beta_z} \end{pmatrix}. \end{aligned} \tag{6}$$

With this input-output system-only channel model, the results from Sec. 3.1 on entangled states now become clear. Firstly, note from Sec. 3.1 that for several system inputs considered, the output state was not a function of $\beta_x$, rather only a function of $\beta_z$. We can now further conclude that the output state is only a function of $\beta_z$ for arbitrary input states. It was also seen in Sec. 3.1 that while with an input of either, an input of $|\uparrow\rangle_{z,S}$ or $|\downarrow\rangle_{z,S}$ leaves the state unchanged and hence acts as an identity map for such input states, in reality there is the possibility of projection to either $|\uparrow\rangle_{z,S}$ or $|\downarrow\rangle_{z,S}$. While not directly affecting $|\uparrow\rangle_{z,S}$ or $|\downarrow\rangle_{z,S}$, such projection does affect the partial density matrix of an entangled state. As the partial density matrix of an entangled state is the maximally mixed state, it is seen that while the maximally mixed state is left unchanged on-average by the Kraus operators in Eqn. (5), the maximally mixed state can be projected to either $|\uparrow\rangle_{z,S}$ or $|\downarrow\rangle_{z,S}$ with respective probabilities $1-\beta$. Such projection, when it occurs, will eliminate the initial two-particle entanglements. This also illustrates that it is important to not only consider on-average effects on the (maximally mixed) density matrix, but also the effects on individual trials. These results are also consistent with the view of the WAY theorem as being a consequence of compatibility of the evolved pointer with the conservation law as passed on to the measured observable [9] and the effect of noise on compatibility of POVMs [10]. It would be of interest for future work to examine if they can be related quantitatively to the present spin model.

**4. Conclusions**

It is generally agreed that momentum and energy are always conserved. Rather than derive the WAY theory using abstract proofs via the detailed properties of unitary evolution and imposing conservation laws, we developed a general class of spin conserving models. The advantage of starting with a conserving model class is that one sees in a straightforward manner the existence of the second and fourth terms in Eqn. (4) which necessarily contribute to an error rate when measuring an eigenstate of $\sigma_x$. This class of models illustrate the general features of the WAY theorem in a manner that is simple and practically accessible.

An input-output system-only model using Kraus superoperators was also found. This showed explicitly that the system is not affected by the correlation in spin angular momentum in the *x*-direction given by $\beta_x$; it is only affected by the correlation $\beta_z$ in the *z*-direction. An interesting question is why does $\beta_x$ drop out. It appears to be related to the fact that the error terms in the second and fourth terms are paired with the same apparatus term $|+\uparrow\rangle_{x,A}$ whilst the terms related to the *z*-component are not the same, i.e. $|+\uparrow\rangle_{z,A}$ appears in one term and $|+\downarrow\rangle_{z,A}$ in another. It is this difference in the *z*-component that prevents cancellation of the second and fourth terms.

Additionally, the three superoperators were shown to be given by two rank-one projection operators into spin $|\uparrow\rangle_{z,S}$ and $|\downarrow\rangle_{z,S}$ weighted by $\sqrt{1-\beta_z}$ and one operator proportional to the identity. This explicitly shows that when $\beta_z$ is not sufficiently high, there will projections into $|\uparrow\rangle_{z,S}$ or $|\downarrow\rangle_{z,S}$ which would cause a Stern-Gerlach device setup to readout spin in the *x*-direction in a noisy manner. This also explains why the entanglement of a maximally entangled state with one spin input into such a Stern-Gerlach device will also be reduced.

**Acknowledgments**

MATLAB Symbolic Toolbox was utilized for deriving equations in Sec. 3. MATLAB was also utilized to double-check the closed-form equations numerically.

## Appendix A

Note that for the models considered in Sec. 3.1 and 3.2, the dimension of the kets $|\downarrow\rangle_{x,S}$ is 2 and $|+\uparrow\rangle_{x,A}$ is 4, and so a typical vector such as $|\downarrow\rangle_{x,S}\otimes|+\uparrow\rangle_{x,A}\otimes|+\uparrow_{\perp}\rangle_{z,A}\otimes|\chi_{\downarrow,x}\rangle_{p,A}$ has dimension of 128.
However, since the last component of the pointer is redundant with the system, it can be removed for computational purposes, reducing the dimension to 32. The unitary matrix utilized in Sec. 3.1 and 3.2 is found as $U =$

$$
\begin{pmatrix}
0 & 0 & 0 & 0 & 0 & \sqrt{\beta_x}\,\sqrt{\beta_z} & \sigma_7 & 0 & 0 & 0 & 0 & 0 & 0 & \sigma_6 & \sigma_5 & 0 & 0 & 0 & 0 & 0 & 0 & 0 & 0 & 0 & 0 & 0 & 0 & 0 & 0 & 0 & 0 & 0 \\
0 & \mathrm{i} & 0 & 0 & 0 & 0 & 0 & 0 & 0 & 0 & 0 & 0 & 0 & 0 & 0 & 0 & 0 & 0 & 0 & 0 & 0 & 0 & 0 & 0 & 0 & 0 & 0 & 0 & 0 & 0 & 0 & 0 \\
0 & 0 & \mathrm{i} & 0 & 0 & 0 & 0 & 0 & 0 & 0 & 0 & 0 & 0 & 0 & 0 & 0 & 0 & 0 & 0 & 0 & 0 & 0 & 0 & 0 & 0 & 0 & 0 & 0 & 0 & 0 & 0 & 0 \\
0 & 0 & 0 & \mathrm{i} & 0 & 0 & 0 & 0 & 0 & 0 & 0 & 0 & 0 & 0 & 0 & 0 & 0 & 0 & 0 & 0 & 0 & 0 & 0 & 0 & 0 & 0 & 0 & 0 & 0 & 0 & 0 & 0 \\
0 & 0 & 0 & 0 & \mathrm{i} & 0 & 0 & 0 & 0 & 0 & 0 & 0 & 0 & 0 & 0 & 0 & 0 & 0 & 0 & 0 & 0 & 0 & 0 & 0 & 0 & 0 & 0 & 0 & 0 & 0 & 0 & 0 \\
\sqrt{\beta_x}\,\sqrt{\beta_z} & 0 & 0 & 0 & 0 & \sigma_{13} & \sigma_3 & 0 & 0 & 0 & 0 & 0 & 0 & \sigma_4 & \sigma_1 & 0 & 0 & 0 & 0 & 0 & 0 & 0 & 0 & 0 & 0 & 0 & 0 & 0 & 0 & 0 & 0 & 0 \\
\sigma_7 & 0 & 0 & 0 & 0 & \sigma_3 & \sigma_{12} & 0 & 0 & 0 & 0 & 0 & 0 & \sigma_1 & \sigma_{10} & 0 & 0 & 0 & 0 & 0 & 0 & 0 & 0 & 0 & 0 & 0 & 0 & 0 & 0 & 0 & 0 & 0 \\
0 & 0 & 0 & 0 & 0 & 0 & 0 & \mathrm{i} & 0 & 0 & 0 & 0 & 0 & 0 & 0 & 0 & 0 & 0 & 0 & 0 & 0 & 0 & 0 & 0 & 0 & 0 & 0 & 0 & 0 & 0 & 0 & 0 \\
0 & 0 & 0 & 0 & 0 & 0 & 0 & 0 & \mathrm{i} & 0 & 0 & 0 & 0 & 0 & 0 & 0 & 0 & 0 & 0 & 0 & 0 & 0 & 0 & 0 & 0 & 0 & 0 & 0 & 0 & 0 & 0 & 0 \\
0 & 0 & 0 & 0 & 0 & 0 & 0 & 0 & 0 & \mathrm{i} & 0 & 0 & 0 & 0 & 0 & 0 & 0 & 0 & 0 & 0 & 0 & 0 & 0 & 0 & 0 & 0 & 0 & 0 & 0 & 0 & 0 & 0 \\
0 & 0 & 0 & 0 & 0 & 0 & 0 & 0 & 0 & 0 & \mathrm{i} & 0 & 0 & 0 & 0 & 0 & 0 & 0 & 0 & 0 & 0 & 0 & 0 & 0 & 0 & 0 & 0 & 0 & 0 & 0 & 0 & 0 \\
0 & 0 & 0 & 0 & 0 & 0 & 0 & 0 & 0 & 0 & 0 & \mathrm{i} & 0 & 0 & 0 & 0 & 0 & 0 & 0 & 0 & 0 & 0 & 0 & 0 & 0 & 0 & 0 & 0 & 0 & 0 & 0 & 0 \\
0 & 0 & 0 & 0 & 0 & 0 & 0 & 0 & 0 & 0 & 0 & 0 & \mathrm{i} & 0 & 0 & 0 & 0 & 0 & 0 & 0 & 0 & 0 & 0 & 0 & 0 & 0 & 0 & 0 & 0 & 0 & 0 & 0 \\
\sigma_6 & 0 & 0 & 0 & 0 & \sigma_4 & \sigma_1 & 0 & 0 & 0 & 0 & 0 & 0 & \sigma_{11} & \sigma_2 & 0 & 0 & 0 & 0 & 0 & 0 & 0 & 0 & 0 & 0 & 0 & 0 & 0 & 0 & 0 & 0 & 0 \\
\sigma_5 & 0 & 0 & 0 & 0 & \sigma_1 & \sigma_8 & 0 & 0 & 0 & 0 & 0 & 0 & \sigma_2 & \sigma_9 & 0 & 0 & 0 & 0 & 0 & 0 & 0 & 0 & 0 & 0 & 0 & 0 & 0 & 0 & 0 & 0 & 0 \\
0 & 0 & 0 & 0 & 0 & 0 & 0 & 0 & 0 & 0 & 0 & 0 & 0 & 0 & 0 & \mathrm{i} & 0 & 0 & 0 & 0 & 0 & 0 & 0 & 0 & 0 & 0 & 0 & 0 & 0 & 0 & 0 & 0 \\
0 & 0 & 0 & 0 & 0 & 0 & 0 & 0 & 0 & 0 & 0 & 0 & 0 & 0 & 0 & 0 & 0 & 0 & 0 & 0 & 0 & \sqrt{\beta_x}\,\sqrt{\beta_z} & 0 & \sigma_7 & 0 & 0 & 0 & 0 & 0 & \sigma_6 & 0 & \sigma_5 \\
0 & 0 & 0 & 0 & 0 & 0 & 0 & 0 & 0 & 0 & 0 & 0 & 0 & 0 & 0 & 0 & 0 & \mathrm{i} & 0 & 0 & 0 & 0 & 0 & 0 & 0 & 0 & 0 & 0 & 0 & 0 & 0 & 0 \\
0 & 0 & 0 & 0 & 0 & 0 & 0 & 0 & 0 & 0 & 0 & 0 & 0 & 0 & 0 & 0 & 0 & 0 & \mathrm{i} & 0 & 0 & 0 & 0 & 0 & 0 & 0 & 0 & 0 & 0 & 0 & 0 & 0 \\
0 & 0 & 0 & 0 & 0 & 0 & 0 & 0 & 0 & 0 & 0 & 0 & 0 & 0 & 0 & 0 & 0 & 0 & 0 & \mathrm{i} & 0 & 0 & 0 & 0 & 0 & 0 & 0 & 0 & 0 & 0 & 0 & 0 \\
0 & 0 & 0 & 0 & 0 & 0 & 0 & 0 & 0 & 0 & 0 & 0 & 0 & 0 & 0 & 0 & 0 & 0 & 0 & 0 & \mathrm{i} & 0 & 0 & 0 & 0 & 0 & 0 & 0 & 0 & 0 & 0 & 0 \\
0 & 0 & 0 & 0 & 0 & 0 & 0 & 0 & 0 & 0 & 0 & 0 & 0 & 0 & 0 & 0 & \sqrt{\beta_x}\,\sqrt{\beta_z} & 0 & 0 & 0 & 0 & \sigma_{13} & 0 & \sigma_3 & 0 & 0 & 0 & 0 & 0 & \sigma_4 & 0 & \sigma_1 \\
0 & 0 & 0 & 0 & 0 & 0 & 0 & 0 & 0 & 0 & 0 & 0 & 0 & 0 & 0 & 0 & 0 & 0 & 0 & 0 & 0 & 0 & \mathrm{i} & 0 & 0 & 0 & 0 & 0 & 0 & 0 & 0 & 0 \\
0 & 0 & 0 & 0 & 0 & 0 & 0 & 0 & 0 & 0 & 0 & 0 & 0 & 0 & 0 & 0 & \sigma_7 & 0 & 0 & 0 & 0 & \sigma_3 & 0 & \sigma_{12} & 0 & 0 & 0 & 0 & 0 & \sigma_1 & 0 & \sigma_{10} \\
0 & 0 & 0 & 0 & 0 & 0 & 0 & 0 & 0 & 0 & 0 & 0 & 0 & 0 & 0 & 0 & 0 & 0 & 0 & 0 & 0 & 0 & 0 & 0 & \mathrm{i} & 0 & 0 & 0 & 0 & 0 & 0 & 0 \\
0 & 0 & 0 & 0 & 0 & 0 & 0 & 0 & 0 & 0 & 0 & 0 & 0 & 0 & 0 & 0 & 0 & 0 & 0 & 0 & 0 & 0 & 0 & 0 & 0 & \mathrm{i} & 0 & 0 & 0 & 0 & 0 & 0 \\
0 & 0 & 0 & 0 & 0 & 0 & 0 & 0 & 0 & 0 & 0 & 0 & 0 & 0 & 0 & 0 & 0 & 0 & 0 & 0 & 0 & 0 & 0 & 0 & 0 & 0 & \mathrm{i} & 0 & 0 & 0 & 0 & 0 \\
0 & 0 & 0 & 0 & 0 & 0 & 0 & 0 & 0 & 0 & 0 & 0 & 0 & 0 & 0 & 0 & 0 & 0 & 0 & 0 & 0 & 0 & 0 & 0 & 0 & 0 & 0 & \mathrm{i} & 0 & 0 & 0 & 0 \\
0 & 0 & 0 & 0 & 0 & 0 & 0 & 0 & 0 & 0 & 0 & 0 & 0 & 0 & 0 & 0 & 0 & 0 & 0 & 0 & 0 & 0 & 0 & 0 & 0 & 0 & 0 & 0 & \mathrm{i} & 0 & 0 & 0 \\
0 & 0 & 0 & 0 & 0 & 0 & 0 & 0 & 0 & 0 & 0 & 0 & 0 & 0 & 0 & 0 & \sigma_6 & 0 & 0 & 0 & 0 & \sigma_4 & 0 & \sigma_1 & 0 & 0 & 0 & 0 & 0 & \sigma_{11} & 0 & \sigma_2 \\
0 & 0 & 0 & 0 & 0 & 0 & 0 & 0 & 0 & 0 & 0 & 0 & 0 & 0 & 0 & 0 & 0 & 0 & 0 & 0 & 0 & 0 & 0 & 0 & 0 & 0 & 0 & 0 & 0 & 0 & \mathrm{i} & 0 \\
0 & 0 & 0 & 0 & 0 & 0 & 0 & 0 & 0 & 0 & 0 & 0 & 0 & 0 & 0 & 0 & \sigma_5 & 0 & 0 & 0 & 0 & \sigma_1 & 0 & \sigma_8 & 0 & 0 & 0 & 0 & 0 & \sigma_2 & 0 & \sigma_9
\end{pmatrix}
$$

where

$\sigma_1 = -\sqrt{\beta_x}\,\sqrt{\beta_z}\,\sqrt{1-\beta_x}\,\sqrt{1-\beta_z}\,\mathrm{i}$

$\sigma_2 = \sqrt{\beta_z}\,(\beta_x - 1)\,\sqrt{1-\beta_z}\,\mathrm{i}$

$\sigma_3 = -\beta_x\,\sqrt{\beta_z}\,\sqrt{1-\beta_z}\,\mathrm{i}$

$\sigma_4 = -\sqrt{\beta_x}\,\beta_z\,\sqrt{1-\beta_x}\,\mathrm{i}$

$\sigma_5 = \sqrt{1-\beta_x}\,\sqrt{1-\beta_z}$

$\sigma_6 = \sqrt{\beta_z}\,\sqrt{1-\beta_x}$

$\sigma_7 = \sqrt{\beta_x}\,\sqrt{1-\beta_z}$

$\sigma_8 = \sqrt{\beta_x}\,\sqrt{1-\beta_x}\,(\beta_z - 1)\,\mathrm{i}$

$\sigma_9 = \beta_x\,\mathrm{i} + \beta_z\,\mathrm{i} - \beta_x\beta_z\,\mathrm{i}$

$\sigma_{10} = \sqrt{\beta_x}\,(\beta_z\,\mathrm{i} - \mathrm{i})\,\sqrt{1-\beta_x}$

$\sigma_{11} = -\beta_z\,\mathrm{i} + \beta_x\beta_z\,\mathrm{i} + \mathrm{i}$

$\sigma_{12} = -\beta_x\,\mathrm{i} + \beta_x\beta_z\,\mathrm{i} + \mathrm{i}$

$\sigma_{13} = -\beta_x\beta_z\,\mathrm{i} + \mathrm{i}$

Note that derivation of several of the results in Sec 3 such as Eqn. (6) that utilize this unitary are rather involved; MATLAB symbolic toolbox was utilized and the results were further double-checked numerically.

[1] E. P. Wigner, "Die Messung quantenmechanischer Operatoren," *Zeitschrift fur Physik (English translation by Paul Busch, https://arxiv.org/abs/1012.4372)*, vol. 133, pp. 101–108, 1952.

[2] H. Araki and M. M. Yanase, "Measurement of quantum mechanical operators," *Physical Review*, vol. 120, p. 622, 1960.

[3] M. M. Yanase, "Optimal measuring apparatus," *Physical Review*, vol. 123, p. 666, 1961.

[4] A. Shimony and H. Stein, "A problem in Hilbert space theory arising from the quantum theory of measurement," *The American Mathematical Monthly*, vol. 86, pp. 292–293, 1979.

[5] G. C. Ghirardi, F. Miglietta, A. Rimini, and T. Weber, "Limitations on quantum measurements. I. Determination of the Minimal Amount of Nonideality and Identification of the Optimal Measuring Apparatuses," *Physical Review D*, vol. 24, no. 2, p. 347, 1981.

[6] G. C. Ghirardi, F. Miglietta, A. Rimini, and T. Weber, "Limitations on quantum measurements. II. Analysis of a model example," *Physical Review D*, vol. 24, no. 2, p. 353, 1981.

[7] M. Ozawa, "Conservation laws, uncertainty relations, and quantum limits of measurements," *Phys. Rev. Lett.*, vol. 88, p. 50402, 2002.

[8] J. Preskill, "Preskill's Notes," http://theory.caltech.edu/~preskill/ph219/chap3_15.pdf.

[9] M. Tukiainen, "Wigner-Araki-Yanase theorem beyond conservation laws," *Phys. Rev. A (Coll. Park).*, vol. 95, no. 1, p. 12127, Jan. 2017, doi: 10.1103/PhysRevA.95.012127.

[10] M. J. Renner, "Compatibility of generalized noisy qubit measurements," *Phys. Rev. Lett.*, vol. 132, no. 25, p. 250202, 2024.